\documentclass[12pt]{article}
\usepackage{latexsym,epsfig,amssymb,euscript,amsmath,verbatim,color,hyperref}

\newcommand{\beq}{\begin{equation}}
\newcommand{\eeq}{\end{equation}}
\newcommand{\beqs}{\begin{eqnarray}}
\newcommand{\eeqs}{\end{eqnarray}}
\newcommand{\bit}{\begin{itemize}}
\newcommand{\eit}{\end{itemize}}
\newcommand{\bce}{\begin{center}}
\newcommand{\ece}{\end{center}}
\newcommand{\ben}{\begin{enumerate}}
\newcommand{\een}{\end{enumerate}}

\newcommand{\hc}{\mathrm{h.c.}}

\newcommand{\nn}{\nonumber}

\newcommand{\GHC}{G_{\mathrm{HC}}}
\newcommand{\GSM}{G_{\mathrm{SM}}}
\newcommand{\NHC}{N_{\mathrm{HC}}}
\newcommand{\GCUS}{G_{\mathrm{cus.}}}

\newcommand{\Fu}{{\mathbf{F}}}
\newcommand{\An}{{\mathbf{A}}}
\newcommand{\Sy}{{\mathbf{S}}}
\newcommand{\Ad}{{\mathbf{Ad}}}
\newcommand{\Sp}{{\mathbf{Spin}}}

\begin{document}

\pagestyle{empty}

\begin{center}

{\LARGE{\bf Fermionic UV completions of \\ \bigskip Composite Higgs models }}

\vspace{1.8cm}

{\large{Gabriele Ferretti$^1$ and Denis Karateev$^2$}}

\vspace{1.5cm}

{\it $^1$ Department of Fundamental Physics, \\
Chalmers University of Technology, 41296 G\"oteborg, Sweden}

\vspace{0.5cm}

{\it $^2$ SISSA, Via Bonomea 265, 34136 Trieste, Italy}

\vspace{2.5cm}

{\bf Abstract}

\vspace{1.cm}

\begin{minipage}[h]{14.0cm}
We classify the four-dimensional purely fermionic gauge theories that give a UV completion of composite Higgs models.
Our analysis is at the group theoretical level, addressing the necessary (but not sufficient) conditions for the viability of these models, such as the existence of top partners and custodial symmetry. The minimal cosets arising are those of type $SU(5)/SO(5)$ and $SU(4)/Sp(4)$. We list all the possible ``hyper-color'' groups allowed and point out the simplest and most promising ones.

\emph{Note added January 2016:} Coset of type $SU(4)\times SU(4)'/SU(4)_D$ added to the classification.

\end{minipage}

\end{center}

\newpage

\setcounter{page}{1} \pagestyle{plain} \renewcommand{\thefootnote}{\arabic{footnote}} \setcounter{footnote}{0}

\section{Introduction}

The discovery of a 126 GeV Higgs boson \cite{LHC} has sharpened the hierarchy problem by confirming the existence of a scalar particle at the electroweak (EW) scale. Broadly speaking, the mass of the Higgs boson can be stabilized against a higher scale in two ways, either by (broken) supersymmetry or by a (broken) shift symmetry. The second case is realized when the Higgs boson appears as a pseudo Nambu-Goldstone boson (pNGB) and is the case of interest for this paper.

The first task of the Higgs boson is to give mass to the vector bosons via BEH mechanism~\cite{BEH}.
This can be accomplished in a strongly coupled theory with a global symmetry group $G_F$ spontaneously broken to a subgroup $H_F$ containing the EW group. Turning on the EW interactions and the top quark couplings turns some of the NGBs in $G_F/H_F$ into pNGBs~\cite{Kaplan:1983fs}. Their vacuum expectation value can then give rise to EW symmetry breaking.

A separate task accomplished by the Higgs boson in the Standard Model (SM) is that of giving mass to the fermions. In this context, it is difficult to obtain large enough top quark masses without fine tuning. A more promising avenue seems to be the introduction of additional fermionic fields from the strong sector coupling linearly to the top (and possibly also to the other fermions) and mediating the EW breaking. This idea was introduced by Kaplan~\cite{Kaplan:1991dc} and goes under the name of partial compositeness. We refer to the review~\cite{Contino:2010rs} for a thorough exposition and a list of relevant references.

Most of the literature so far has concentrated on the phenomenological aspects of these models, assuming that the low energy lagrangian has such properties and using the CCWZ formalism~\cite{CCWZ} without asking for a UV description\footnote{We are considering here only four-dimensional models. Much work has been done in higher dimensional models and we refer again to~\cite{Contino:2010rs} for the relevant literature.}. Notable exceptions are the work~\cite{Caracciolo:2012je} in the context of supersymmetric theories and very recently~\cite{Barnard:2013zea} in the context of strongly coupled gauge theory with a purely fermionic matter content. Similar thoughts had been pursued by us and presented in~\cite{talk}.

The purpose of this short note is to classify the possible UV completions that are available for four-dimensional models of partial compositeness with a purely fermionic matter content. We will not explore here the more phenomenological constraints arising from the dynamics, but we feel that this classification is useful to get an idea of the possibilities available. Needless to say, our classification will only be as good as the assumptions we make.

The paper is organized as follows. In Section~2 we present the ``wish list'' of conditions that we want our theory to obey. These conditions are divided into assumptions that we make to concretely define the framework in which we work and consistency requirements that are necessary (but by no means sufficient!) for these theories to be acceptable models of partial compositeness. In Section~3 we present the solutions to our requirements. In Section~4 we take a more critical look at the models obtained and discuss which ones seem more promising for phenomenological applications.

\section{The wish list}

We begin in \ref{assuptions} by spelling out the framework in which we work. Within this framework, the model has to satisfy certain simple conditions in order to be a viable candidate for partial compositeness. We list these conditions in~\ref{conditions}.

\subsection{Assumptions}
\label{assuptions}
We look for a microscopic theory of partial compositeness based on a \emph{simple} hyper-color group $\GHC$ and \emph{only} (LH Weyl) fermions $\psi \in n_1 R_1 + \dots + n_p R_p$. $R_i$ is an irreducible representation (irrep) of $\GHC$ repeated $n_i$ times. For $i \neq j$ it is always intended $R_i \neq R_j$. Thus $p$ is the number of different irreps in the model. (The model of~\cite{Barnard:2013zea} corresponds to $p=2$.)

Restricting the search to simple hyper-color groups is motivated by a criterion of minimality. In the absence of additional discrete symmetries one expects the different simple factors to become strongly coupled at different scales and one could restrict the attention to the last step. Discrete symmetries could be used to keep the gauge couplings of different groups to be equal but we will not make this extra assumption.

This choice already implies that the anomaly-free global symmetry group is $ G_F = SU(n_1)\times \dots \times SU(n_p) \times U(1)^{p-1}$.
The SM fermions are neutral under $\GHC$ whereas the hyper-fermions $\psi$ will be given appropriate charge under $\GSM$ to have the appropriate bound states serving as composite Higgs and top partners.

We will limit our search to asymptotically free (AF) theories. It may seem that the requirement of asymptotic freedom is not relevant because we are going to need some four-fermi interactions to couple the elementary top quark to its composite partners. What is crucial however is that the theory becomes strongly coupled in the IR and this is generically attained in a AF theory if one starts at small coupling in the UV and there are no additional perturbative fixed points. There could be other phases, but we will not consider that possibility. We recall for convenience that the $\beta$-function is $\beta(\alpha_{\mathrm{HC}}) = - b_0 \;\alpha^2_{\mathrm{HC}}/(24\pi) + \dots $, with
$ b_0 =  11 C(\Ad) - 2 \sum_R  T(R)$, where the sum is over all irreps (for  Weyl fermions) including the degeneracy.

One further step, that will not be investigated here, is to try to generate the required four-fermi couplings by including both $\GSM$ and $\GHC$ into an extended hyper-color theory. This seems like a tall order but as a first step one should have a list of possible theories that can be merged anyway.

\subsection{Consistency conditions}
\label{conditions}
The basic requirements we impose in order for our model to be a potential candidate are
\ben
\item Absence of gauge anomalies for $\GHC = SU(N)$ and global anomalies for  $\GHC = Sp(2N)$.
\item The \emph{possibility} of a symmetry breaking pattern: $G_F \rightarrow H_F \supset \GCUS \supset \GSM$, where we defined $\GCUS =
SU(3)_c\times SU(2)_L\times SU(2)_R\times U(1)_X$ and, of course, $\GSM = SU(3)_c\times SU(2)_L\times U(1)_Y$.
\item $\GSM$ free of 't Hooft anomalies.
\item $G_F/H_F \ni ({\mathbf{1}},{\mathbf{2}},{\mathbf{2}})_0$ of $\GCUS$.
\item $\psi^3$ hyper-color singlets that can be used as top partners. The minimal requirement being that they are spinors whose LH components have $\GSM$ quantum numbers opposite to the third generation of quarks $Q_L$ and $t^c_R$, namely $(\bar{\mathbf{3}}, \mathbf{2})_{-1/6}$, $(\mathbf{3}, \mathbf{1})_{2/3}$.
\een

We spend a few words on some of these condition.

The requirement that $H_F$ contains $\GCUS$ and not simply $\GSM$ follows from the requirement of custodial symmetry. An extra $SU(2)_R$ is needed to avoid large tree-level corrections to the $\rho$ parameter. However, if the SM hyper-charge $Y$ were to be obtained from $SU(2)_R$ only, one could not get realistic values for the fermion fields. Hence the necessity of introducing the extra $U(1)_X$. In these model: $Y = T_R^3 + X$.

When investigating the possible symmetry breaking patterns we assume that at strong coupling a condensate forms. In the case where this condensates affects one single $SU(n)$ flavor group, symmetry reasons and the analysis of~\cite{Preskill:1981sr,VW} implies that the only possibilities are for $SU(n)$  to break to $SO(n)$ (if the fermionic bilinear is formed with fermions in a real irrep $R$) or to
$Sp(n)$ (if $R$ is pseudo-real). This is easily understood by recalling that a real (pseudo-real) irrep has a symmetric (anti-symmetric) invariant two-tensor of $\GHC$ and that this determines the corresponding symmetry properties of the order parameter. We will return to this issue and its relation to the maximally attractive channel (MAC) hypothesis~\cite{Raby:1979my} in Section~4.

The subgroup $\GSM$ of $H_F$ must be free of 't Hooft anomalies from the obvious fact that we need to gauge it when coupling the hyper-color fermions $\psi$ to the SM.

One of the NGB $H$ in $G_F/H_F$ must transform as the $({\mathbf{1}},{\mathbf{2}},{\mathbf{2}})_0$ of $\GCUS$ in order to be the candidate for the composite Higgs. In general there will be other pNGB in the model. They should obviously acquire a mass that is consistent with the current bounds. The field $H$ should be ``misaligned'', presumably by the coupling with the top quark, and condense. This phenomena should occur along the same lines as in the effective field theory language reviewed in~\cite{Contino:2010rs} and we will not discuss it.

Lastly, we must be able to form fermionic bound states of type $\psi^3$ that can be interpreted as top quark partners. The existence of a linear coupling requires that they are spinors with the same $\GSM$ quantum numbers as the third generation of quarks $Q_L$ and $t_R$. We will see that various possible quantum numbers with respect to $H_F$ are possible. We will not impose the existence of partners for each SM fermion, the top quark being the most pressing issue. Also, in order for this to work, the trilinear in question must pick up a large anomalous dimension, taking them from the perturbative value $9/2$ to near $5/2$. That this is possible in some models has been argued in \cite{Barnard:2013zea} but we will not address this issue in this note. If there are fermions $\lambda$ in the adjoint representation, there is also the possibility of constructing dimension $7/2$ fermionic invariants\footnote{We learned from M.~Serone that this possibility has been suggested by A.~Wulzer.} of the type $F_{\mu\nu}^a\sigma^\mu\bar\sigma^\nu\lambda^a$.

We do not require the existence of partners for every SM fermion. In fact, in many of the models we find, this is not possible if the search is restricted to trilinear $\GHC$ invariants only. We return to this issue in the conclusions.

\section{Solution to the constraints}

We will now classify the models satisfying the above requirements for the case of $p=1, 2$~and~$3$. The cases $p=1$~and~$3$ were presented in~\cite{talk}\footnote{\label{errata} The $p=3$ case presented in~\cite{talk} was actually incomplete. Here we present the full list of models in Table~\ref{p3}.}. At the time, the case $p=2$ did not seem promising, but the appearance of the model~\cite{Barnard:2013zea} made us reconsider this case and we now include it in the discussion. Models with $p>3$ do not seem to add any substantial new feature and we will not discuss them.

Given a $\GHC$ and a set of $R_i$ satisfying the constraints, for each $R_i$ there will be generically a range of integers $n_i$ allowed. We will list only the smallest possible values of $n_i$ in each case. Within each class of solutions one can easily find the other values of $n_i$ allowed by checking when one loses asymptotic freedom. This leads to non-minimal models with a larger set of fields and bigger cosets. Good sources for the group theoretical material required in some of the calculations are~\cite{LIE}. In the following, we will denote specific irreps either by their dimensionality or by the symbols $\Fu,~\Sy_n,~\An_n,~\Ad$ and $\Sp$ for the fundamental, $n-$symmetric, $n-$antisymmetric, adjoint and spin.

\subsection{The  $p=1$ case}

This is a somewhat unrealistic case but we discuss it for completeness and to illustrate the point.
Consider $ \psi \in R$ a unique representation repeated $n$ times yielding $G_F = SU(n)$.
We need to break $G_F$ with a gauge invariant bilinear condensate $ \langle\psi^2\rangle$ and this can be formed only if
the irrep $R$ is real or pseudoreal.

The further requirement that there be $\psi^3$ invariants forces $R$ to be real and the required symmetry breaking pattern becomes $SU(n)\to SO(n)$.
Assuming that this is the case, $\GCUS \subset SO(n)$ requires $n \geq 10$.
However $({\mathbf{1}},{\mathbf{2}},{\mathbf{2}})_0 \not\in SU(10)/SO(10)$ which requires instead $ n \geq 11$.
This last point can be seen by showing that the decomposition of the $\Sy_2 = {\mathbf{54}}$ of $SO(10)$ contains no Higgs candidates. (The $X$-charge is arbitrarily normalized here.)
\beqs
   {\mathbf{54}} &\rightarrow &\phantom{+}({\mathbf{8}},{\mathbf{1}},{\mathbf{1}})_0 + ({\mathbf{6}},{\mathbf{1}},{\mathbf{1}})_4 +
   (\bar {\mathbf{6}},{\mathbf{1}},{\mathbf{1}})_{-4} +({\mathbf{1}},{\mathbf{3}},{\mathbf{3}})_0
   \\&& + ({\mathbf{3}},{\mathbf{2}},{\mathbf{2}})_{-2} + (\bar {\mathbf{3}},{\mathbf{2}},{\mathbf{2}})_2 + ({\mathbf{1}},{\mathbf{1}},{\mathbf{1}})_0 \nn \label{deco54}
\eeqs
On the other hand, already with $n=11$ there will be one such field as can be seen by recalling the decomposition of the $\Sy_2$ of $SO(11)$ into $SO(10)$:
$\mathbf{65}\to \mathbf{54} + \mathbf{10} + \mathbf{1}$ and that
\beq
   \mathbf{10}\rightarrow ({\mathbf{1}},{\mathbf{2}},{\mathbf{2}})_0 + ({\mathbf{3}},{\mathbf{1}},{\mathbf{1}})_{-2} +
   (\bar{\mathbf{3}},{\mathbf{1}},{\mathbf{1}})_{2} \label{deco10}
\eeq

Asymptotic freedom requires thus $ T(R) < \frac{1}{2}T(\Ad)$.

This already rules out many possible choices for $\GHC$. The further top partner requirement singles out $ G_2$ (with $ R=\mathbf{7}$)  and $ F_4$ (with $ R=\mathbf{26}$) as the only possibilities\footnote{Fun fact: The 't Hooft anomaly matching condition cannot be satisfied for $ G_2$ which implies that $ G_F$ must be broken.}. Embedding the group $\GCUS$ into $SO(11)$ as above one can obtain the correct quantum numbers for the top partners.

We do not believe these models to be promising because of the difficulties with proton stability. Amongst the large number of pNGB there will be some that mediate proton decay, confirming assertions made in e.g.~\cite{oai:arXiv.org:0910.1789}. Even if one strictly couples only the top quark to the composite sector, one expects $B$ violating terms to be induced at higher orders unless a symmetry is enforced to prevent this.
Attempts at saving models of this ``GUT''-type are made in~\cite{Frigerio:2011zg} albeit with a different coset.
For instance, one could try $ n>11$ and imposing an \emph{ad hoc} symmetry that commutes with the whole $\GSM$ but this would require many repetitions and incomplete (split) multiplets.

\subsection{The  $p=2$ case}

In this case we assume $\psi \in m R_1 + n R_2$, yielding a flavor group $G_F = SU(m)\times SU(n)\times U(1)$. We want to be allowed to break the first $SU(m)$ to a custodial subgroup containing $SU(2)_L\times SU(2)_R$ and the Higgs in the appropriate representation.
This means that $m \geq 4$ if $R_1$ is pseudo-real (coset $SU(4)/Sp(4)$) or $m\geq 5 $ if $R_1$ is real (coset $SU(5)/SO(5)$)\footnote{The smallest cosets of this type have been classified in \cite{Mrazek:2011iu}. Models of the type $SU(5)/SO(5)$ have been investigated in~\cite{Vecchi:2013bja}. Models of the type $SU(4)/Sp(4)\equiv SO(6)/SO(5)$ have been investigated in~\cite{Gripaios:2009pe}. For earlier investigations of $SU(4)$ cosets in the context of technicolor, see \cite{su4}.}.

The other $SU(n)$ group is required to contain the group $SU(3)_c\times U(1)_X$ in an anomaly free way. This requires $n\geq 6$ and we will assume in the discussion the minimal case $n=6$. The fermions can be decomposed as in Table~\ref{su3emb}.
Note that in principle $SU(6)$ could be broken to one of its special maximal subgroups $SO(6)\equiv SU(4)$ or $Sp(6)$ and that would still allow one to embed $SU(3)_c\times U(1)_X$ in an anomaly free way as we have essentially done in the $p=1$ case. This case would give rise to additional pNGB's that may be lifted if an $SU(3)_c\times U(1)_X$ invariant mass can be constructed. We shall return to this issue in Section~4.
\begin{table}
  \centering
  \begin{tabular}{c|c|c|c|}
      & $\GHC$ & $SU(m=4~\mathrm{ or }~5)$ & $SU(6)\supset SU(3)_c\times U(1)_X$ \\
      \hline
    $\psi_1 \equiv \psi$ & $R_1$ & $\Fu$ & $\mathbf{1}_0$ \\
    $\psi_2 = \left(
                \begin{array}{c}
                  \chi\\
                  \tilde\chi\\
                \end{array}
              \right) $    & $R_2$ & $\mathbf{1}$ & $\mathbf{6} = \left(
                                                                    \begin{array}{c}
                                                                     {\mathbf{3}}_X \\
                                                                      {\bar{\mathbf{3}}}_{-X} \\
                                                                    \end{array}
                                                                  \right)$  \\
    \hline
  \end{tabular}
  \caption{The fermions of the UV theory and their quantum numbers. The charge $X$ will be equal to $2/3$ in those cases were the top partners are constructed out of one $\psi_2$ field and equal to $-1/3$ in the cases were one needs two such fields. To avoid cluttering we do not write this charge in the text. }\label{su3emb}
\end{table}

We can immediately eliminate the possibility $\GHC = SU(\NHC)$.
In this case $b_0 = 22\NHC - 2 m T(R_1) - 12 T(R_2) $, where $m=4,(5)$ depending on $R_1$ pseudo-real (real) and we set $n=6$ from the onset.
Since $R_1$ is real or pseudo-real, it does not contribute to the gauge anomaly. Thus $R_2$ must be real or pseudo-real as well, since its anomaly could not cancel against anything. The possible irreps with the smallest indices are thus the $\Ad$ (real) or the $\An_n$ (real/pseudo-real for $SU(2n)$, $n$ even/odd). In all such cases one immediately sees that asymptotic freedom is lost. In a similar way one eliminates all the exceptional groups.

For the remaining groups there are interesting solutions, including of course the one in~\cite{Barnard:2013zea}. Let us thus start with $Sp(2\NHC)$. ($\NHC \geq 2$ since $Sp(2) \equiv SU(2)$.) The indices for these groups can be ordered in increasing order as  $T(\Fu), T(\An_2), T(\Ad), \cdots$ ($\Ad = \Sy_2$) with the only exception of $Sp(6)$, in which $T(\An_3) = 5 $ fits in between $T(\An_2)=4$ and $T(\Ad)=8$.
For these groups $T(\An_2)=2\NHC - 2$ and $T(\Ad)=2\NHC +2$. Also, $\Ad$ and $\An_2$ are real, whereas $\Fu$ and $\An_3$ are pseudo-real.

One easily checks that the only case where $\An_3$ is allowed to appear by asymptotic freedom is $Sp(6)$, with four $R_1=\An_3$ and six $R_2=\Fu$. However, since both are pseudoreal, no baryons are allowed and the $\psi^3$ requirement is not fulfilled. The only case where the $\Ad$ can be present without losing asymptotic freedom is five $R_1=\Ad$ and six $\Fu$. (Recall that we are only considering the minimal values of $m$ and $n$ allowed.) In this case there are cubic invariants that can function as top partners of the type $\chi\psi\chi$, $\tilde\chi\psi\tilde\chi$, $\chi^\dagger\psi^\dagger\tilde\chi$, $\tilde\chi^\dagger\psi^\dagger\chi$ and their RH conjugates. (Use ${\mathbf 3}\times{\mathbf 3} = \bar{\mathbf 3}+{\mathbf 6}$ and ${\mathbf 3}\times \bar{\mathbf 3} = {\mathbf 1}+{\mathbf 8}$.). Asymptotic freedom requires $2\NHC \geq 12$.

Similarly, one can replace the five $\psi\in \Ad$ with five $\psi\in\An_2$. The possible top partners now also include two combinations that were not allowed before: $\chi^\dagger\psi\chi^\dagger$ and $\tilde\chi^\dagger\psi\tilde\chi^\dagger$ and their RH conjugates.

The last class of models based on $Sp(2\NHC)$ are the ones discussed in~\cite{Barnard:2013zea}, namely four $R_1=\Fu$ and six $R_2=\An_2$. These models are asymptotically free for $2\NHC \leq 36$. The baryons are now formed with only one of the $\chi$'s and transform in various irreps of $SU(4)$, e.g. ${\mathbf 6}$ as discussed in~\cite{Barnard:2013zea}. By contrast, the previous cases had top partners in the ${\mathbf 5}$ of $SU(5)$. There are no global anomalies in any of these cases.

Lastly, consider $SO(\NHC)$ ($\NHC \geq 7$).
The first two classes are a repetition of the $Sp(2\NHC)$ case with the difference that now $\Ad=\An_2$. Hence we have the case of five $R_1=\Sy_2$ and six $R_2=\Fu$ (with $  \NHC \geq 55$) and five $R_1=\Ad$ and six $R_2=\Fu$ (with $  \NHC \geq 15$). Particularly for the first case, the size of the hyper-color group is too large to be considered interesting for phenomenology, but we include these cases for completeness.

In the case of  $SO(\NHC)$ we must also consider spinor irreps. If both $R_1$ and $R_2$ were spinors, one could not construct top partners. One of the two must be a vector irrep. Since we use only one spinor irrep, we can simply denote it as $\Sp$ without specifying its chirality, the opposite choice being equivalent.  Asymptotic freedom allows only $\Fu$ to appear and we have thus two separate classes: $R_1=\Fu$, $R_2 = \Sp$ and viceversa. In both cases\footnote{Note that the $\Sp$ irreps of $SO(15)$ and $SO(16)$ are both real, so they require $m=5$ and thus cannot be used by the requirement of asymptotic freedom.} we must have $\NHC \leq 14$. There are further restrictions excluding the case $\NHC=8, 12$ for which it is not possible to have top partners.

We summarize the results of this section by listing the models that pass the constraints of Section~2 in Table~\ref{p2}.
\begin{table}
  \centering
  \begin{tabular}{|c|c|c|c|}
    \hline
    $\GHC$ & $R_1$ & $R_2$ & Restrictions \\
    \hline\hline
    $Sp(2 \NHC)$ & $5\times \Ad$ & $6\times \Fu$ & $2 \NHC \geq 12$ \\
    \hline
    $Sp(2 \NHC)$ & $5\times \An_2$ & $6\times \Fu$ & $2 \NHC \geq 4$ \\
    \hline
    $Sp(2 \NHC)$ & $4\times \Fu$ & $6\times \An_2$ & $2 \NHC \leq 36$ \\
    \hline
    $SO(\NHC)$ & $5\times \Sy_2$ & $6\times \Fu$ & $\NHC \geq 55$ \\
    \hline
    $SO(\NHC)$ & $5\times \Ad$ & $6\times \Fu$ & $\NHC \geq 15$ \\
    \hline
    $SO(\NHC)$ & $5\times \Fu$ & $6\times \Sp$ & $\NHC=7,9,10,11,13,14$ \\
    \hline
    $SO(\NHC)$ & $5\times \Sp$ & $6\times \Fu$ & $\NHC = 7,9 $ \\
    \hline
    $SO(\NHC)$ & $4\times \Sp$ & $6\times \Fu$ & $\NHC = 11,13$ \\
    \hline
  \end{tabular}
  \caption{All allowed cases for $p=2$.}\label{p2}
\end{table}

\subsection{The  $p=3$ case}

We move on to the case of three different irreps of $\GHC$ that was presented in~\cite{talk}. (See however footnote~\ref{errata}.)

The main difference from the previous case is that now the QCD group $SU(3)_c$ is embedded as the diagonal group of a semi-simple group $SU(3)\times SU(3)'$ and not $SU(6)$ as before.
The new attempt is thus $ \psi \in m R_1 + 3 R_2 + 3 R_3$ where $ R_i$ are different irreps of some hyper-color group $\GHC$.
This leads to $ G_F = SU(m)\times SU(3)\times SU(3)'\times U(1) \times U(1)'$, with $m=5(4)$ for $R_1$ real(pseudo-real) as before.
\begin{table}
  \centering
  \begin{tabular}{c|c|c|c|}
      & $\GHC$ & $SU(m=4~\mathrm{ or }~5)$ & $SU(3)_c \times U(1)_X$  \\
      \hline
    $\psi_1 \equiv \psi$ & $R_1$ & $\Fu$ & $\mathbf{1}_0$ \\
    $\psi_2 = \chi $   & $R_2$ & $\mathbf{1}$ & ${\mathbf{3}}_{-1/3}$  \\
    $\psi_3 = \tilde\chi $   & $R_3$ & $\mathbf{1}$ & ${\bar{\mathbf{3}}}_{1/3}$  \\
    \hline
  \end{tabular}
  \caption{The fermions of the UV theory for $p=3$ and their quantum numbers. As before, we do not indicate the $X$-charge explicitly in the text.}\label{scheme3}
\end{table}

A further restriction is that $ d(R_2) = d(R_3)$ and $ T(R_2) = T(R_3)$ to allow for the possible symmetric embedding of $SU(3)_c$ to the anomaly free diagonal.
This restriction rules out all symplectic and exceptional groups.
As far as $\GHC=SU(\NHC)$ goes, imposing asymptotic freedom, we can only have $R_1 = \Ad$ ($m=5$) and $R_2=\Fu$, $R_3=\bar\Fu$ and three exceptional cases:
\bit
   \item $\GHC=SU(4)$ with $R_1 = \An_2$ ($m=5$) and $R_2=\Fu$, $R_3=\bar\Fu$
   \item $\GHC=SU(6)$ with $R_1 = \An_3$ ($m=4$) and $R_2=\Fu$, $R_3=\bar\Fu$
   \item $\GHC=SU(6)$ with $R_1 = \An_3$ ($m=4$) and $R_2=\An_2$, $R_3=\bar\An_2$
\eit
Only the first of the three exceptional cases cases is acceptable. The general $SU(\NHC)$ case leads to $\GHC$ singlets in the $\mathbf 8$ of $SU(3)_c$, the $\GHC=SU(6)$ case with $R_2=\Fu$, $R_3=\bar\Fu$ has no baryons whatsoever and the case with $R_2=\An_2$, $R_3=\bar\An_2$ has colored $\GHC$ singlet but none of them carries a non-trivial irrep of $SU(4)$ needed for a top partner. For the acceptable $SU(4)$ case, the top quark partners are $\chi\psi\chi$, $\tilde\chi\psi\tilde\chi$, $\chi^\dagger\psi^\dagger\tilde\chi$, $\tilde\chi^\dagger\psi^\dagger\chi$, $\chi^\dagger\psi\chi^\dagger$, $\tilde\chi^\dagger\psi\tilde\chi^\dagger$ and their RH conjugates.

Consider now $SO(\NHC)$ groups with $\NHC$ even and with $R_2$ and $R_3$ spinor irreps of opposite chirality.
Asymptotic freedom allows $\NHC = 8,10,12,14$ and one can construct top partners for all of these cases.
For the cases $SO(8)$ and $SO(12)$ the partners are restricted to $\chi\psi^\dagger\tilde\chi^\dagger$, $\tilde\chi\psi^\dagger\chi^\dagger$ and their RH conjugates only.
We summarize the results of this section by listing the models that pass the constraints of section~2 in Table~\ref{p3}.
($\Sp$ and $\Sp'$ denote spinor irreps of opposite chirality.) In the case of $SO(8)$ one can use triality to obtain equivalent solutions.
\begin{table}
  \centering
  \begin{tabular}{|c|c|c|c|c|}
    \hline
    $\GHC$ & $R_1$ & $R_2$ & $R_3$ & Restrictions \\
    \hline\hline
    $SU(\NHC)$ & $5\times \An_2$ & $3 \times \Fu $ & $3 \times \bar{\Fu}$ & $\NHC=4$ \\
    \hline
    $SO(\NHC)$ & $5 \times \Fu$ & $3 \times \Sp $ & $3 \times \Sp'$ & $\NHC=8,10,12,14$\\
    \hline
  \end{tabular}
  \caption{All allowed cases for $p=3$. }\label{p3}
\end{table}

\section{Discussion}

In the previous section we classified the purely fermionic gauge theories that give a UV completion of composite Higgs models. The list of models is given in Tables~\ref{p2} and~\ref{p3}. For each class we have given the realization with the minimal set of fields. We listed the models according to the numbers of distinct irreps of $\GHC$ needed but, as far as the IR properties are concerned, a more relevant distinction is between models that yield the coset $SU(5)/SO(5)$ and models that yield $SU(4)/Sp(4)$. The second type is more rare and consists of the model presented in~\cite{Barnard:2013zea} and two more based on $\GHC=SO(11)$ and $SO(13)$.

We have not proven that the symmetry breaking actually occurs, although the analysis of~\cite{Barnard:2013zea} suggests that this is indeed the case for their model and one could easily generalize their argument to any of the models presented here. Given that we only consider asymptotically free theories, it is quite reasonable to expect a bilinear condensate to form as the theory flows to strong coupling in the IR. Given the nature of the irreps and assuming that non-chiral flavor groups are left unbroken we are thus led to the symmetry breaking patterns presented.

Some models comprise an unreasonably large number of fermions that will lead to a Landau pole too close to the EW scale. Amongst the models that allow for a low dimensional $\GHC$ we note the second and third entry in Table~\ref{p2} and the $SU(4)$ model in Table~\ref{p3}.

There is an important refinement of the above discussion that requires some extra care.
According to the heuristics of the MAC hypothesis~\cite{Raby:1979my}, the symmetry breaking patterns occur in a specific order determined by the quadratic Casimir operators $C$ of the various irreps. In short, amongst all irreps $R_i$ one considers the scalar bilinears $R_i\times R_j = \sum_k R'_k$ (including those that might break the $\GHC$ symmetry). The channel that is expected to condense first is the maximally attractive one, i.e. the one with the lowest value of $C(R'_k)-C(R_i)-C(R_j)$. After that, the fermions responsible for the condensate are removed and the process continues.

It would be desirable if this trend was not in conflict with our necessity to keep $\GHC$ and $SU(3)_c$ unbroken at low energies. This condition removes the $SO(10)$ and $SO(14)$ cases from Table~\ref{p2} and the $SO(8)$ and $SO(12)$ cases from Table~\ref{p3}. Consider for example the case $SO(14)$ of Table~\ref{p2} (sixth row). The MAC is $\Sp \times \Sp \to \Fu$, for which $13/2-91/8-91/8=-65/4$, which is even stronger than $\Fu\times\Fu\to{\mathbf 1}$, for which $0-13/2-13/2 =-13$. The strong dynamics would then tend to break the hyper-color group and no mass term can be introduced to prevent this. It seems unlikely that this models give rise to acceptable IR behavior.

The above condition is related to asking for the possibility to write a $\GHC \times SU(3)_c\times U(1)_X$ invariant mass term for the fermions $\chi, \tilde\chi$. The actual form of this mass matrix depends on the symmetry breaking pattern favored by the MAC hypothesis. Consider the $p=2$ case. For those models in which $R_2$ is pseudo-real (the first two rows and the $\NHC=11, 13$ cases in the sixth row in Table~\ref{p2}) one expects a $Sp(6)$ preserving condensate and a suitable mass term can be, up to a global rotation
\beq
    \frac{m}{2} \psi_2^T {\footnotesize{\left(
                 \begin{array}{cc}
                   0 & 1 \\
                   -1 & 0 \\
                 \end{array}
               \right)}}
               \psi_2 + \hc = m \chi \tilde\chi+ \hc
\eeq
having set $\chi=(\psi_2^1, \psi_2^2, \psi_2^3)^T$ and $\tilde\chi=(\psi_2^4, \psi_2^5, \psi_2^6)$ (Weyl and $\GHC$ indices suppressed).
For the remaining models in Table~\ref{p2} the preferred pattern is to a $SO(6)$ condensate. A mass term that preserves this symmetry is
\beq
        m \psi_2^T \psi_2 + \hc = m \chi \tilde\chi+ \hc
\eeq
where now $\chi=(\psi_2^1 + i \psi_2^4, \psi_2^2 + i \psi_2^5, \psi_2^3 + i \psi_2^6)^T$ and \\
$\tilde\chi=(\psi_2^1 - i \psi_2^4, \psi_2^2 - i \psi_2^5, \psi_2^3 - i \psi_2^6)$.

Concerning the remaining cases in Table~\ref{p2} involving spinorial irreps, we also notice that the channel $\Sp\times\Sp\to \mathbf{1}$ becomes more attractive than $\Fu\times\Fu\to \mathbf{1}$ for $\NHC\geq 9$. Thus, for the models in the sixth row, $\NHC=7$ is the only case in which $R_1$ is expected to condense first in the absence of masses and the opposite is true for the models in the last two rows.

In the $p=3$ case, the symmetry breaking goes directly to the diagonal $SU(3)_c$ subgroup and the mass term is $m \chi \tilde\chi + \hc$ from the start. This is allowed in the $SU(4)$, $SO(10)$ and $SO(14)$ cases but not for the $SO(8)$ and $SO(12)$ cases. Also note that, in the $SO(12)$ case, the MAC are $\Sp\times \Sp \to {\mathbf 1}$ and $\Sp'\times \Sp' \to {\mathbf 1}$, tending to break $SU(3)_c$.

An even stronger condition would be to require that no hyper-mesons of any type (even spin one) in the $\mathbf 3$ or $\mathbf 6$ of $SU(3)_c$ be present. Bound states of this type are potentially dangerous because quantum corrections might induce trilinear B-violating couplings between them and two SM fermions. This disfavors all the $p=2$ models for which such mesons can always be formed e.g. as combinations $\chi\chi$ or $\tilde\chi\chi^\dagger$ and singles out the coset $SU(5)/SO(5)$ (the only one attainable in $p=3$ models) and $\GHC=SU(4)$ as its minimal UV realization.

This last case is an interesting model, yielding top partners in the $\mathbf 5$ of $SO(5)$ that decompose, after EW symmetry breaking, into the usual partners plus an exotic top of electric charge $5/3$. The latest bounds on such objects are discussed in~\cite{Chatrchyan:2013wfa}. The pNGB spectrum is obtained by decomposing the $\mathbf{14}$ of $SO(5)$ into $SU(2)_L\times U(1)_Y$
\beq
    {\mathbf{14}} \to {\mathbf 3}_{\pm 1} + {\mathbf 3}_{0} + {\mathbf 2}_{\pm 1/2} + {\mathbf 1}_{0}
\eeq
yielding, assuming that EW symmetry breaking is driven by the doublet, the usual Higgs boson $h$ plus a double-charge meson $\phi^{\pm\pm}$, two single-charge ones $\phi^{\pm}$, $\phi^{\prime \pm}$ and four neutral ones $\phi^0$, $\phi^{\prime 0}$, $\phi^{\prime\prime 0}$, $\eta^0$, the last one being totally neutral under $\GCUS$.

Cosets of type $SU(5)/SO(5)$ are not free from phenomenological problems, such as the couplings of the weak isotriplet, but these problems have been argued~\cite{Vecchi:2013bja} not to be insurmountable. That work also shows that EW breaking is expected to proceed as desired. The extra pNGBs would then be a generic prediction of these models.

Let us also comment on the issue of finding partners to the other SM fermions.
None of the models give rise to cubic composite partners for all states.
Consider for instance the $b^c_R$ and the lighter members of this family. With the exception of the third and last two entries in Table~\ref{p2}, in all other models the $X$-charge of $\chi/\tilde\chi$ must be equal to $\pm 1/3$. This precludes the possibility of having a $b^c_R$ partner amongst the trilinear hyper-color invariants as can be easily checked recalling that such invariants must contain two fields of type $\chi$ or $\tilde\chi$. The remaining models suffer of similar problems in the lepton sector. For instance, it is clearly not possible to construct leptonic partners that are doublets under $SU(2)_L$ with only cubic invariants.

Our view on this issue is that the top quark should be treated differently since it has a mass at the EW scale, compared to all the others masses that, while differing from each other by many orders of magnitude, are all small compared to that scale. For lighter fermions, a more standard quadratic coupling as in extended technicolor could still be viable, although the issue deserves a more careful investigation.

Lastly, given the presence of additional anomaly free $U(1)$s, all of the models presented here also give rise to NGBs that are totally neutral under $\GCUS$ and do not acquire mass under this approximation. We did not discuss them here but their properties would have to be addressed in a cosmological contest.

\subsection*{Acknowledgments}

We thank A.~Romanino and M.~Serone for discussion. M.~Serone has been particularly helpful in getting us started thinking about composite Higgs models. The research of G.F. is supported in part by the Swedish Research Council (Vetenskapsr{\aa}det) contract B0508101.
D.K. wants to thank the THEP Department of Lund University where he began working on this project.
\vspace{-.2cm}

\subsection*{{Note added January 2016:}}

It is also possible to include EW-coset of ``QCD''-type, arising from vector-like fermions in a complex representation of $\GHC$. 
The minimal custodial case is $SU(4)\times SU(4)'/SU(4)_D$. The first case $\big(SU(4)\times SU(4)'/SU(4)_D\big)\times \big(SU(6)/SO(6)\big)$ could logically be added to the $p=3$ list while the second one: $\big(SU(4)\times SU(4)'/SU(4)_D \big)\times \big(SU(3)\times SU(3)'/SU(3)_D\big)$ represents a $p=4$ case. The remaining case $\big(SU(4)\times SU(4)'/SU(4)_D \big)\times \big(SU(6)/Sp(6)\big)$ does not allow for the construction of top partners.

These additional solutions have been presented at~\cite{later} and are reported below in Table~\ref{added}.

\vspace{-.2cm}

\begin{table}[h]
\bce
{\footnotesize
\beq
           \frac{G_{\mathrm{F}}}{H_{\mathrm{F}}} = { \frac{SU(4)\times SU(4)'}{SU(4)_D}}{ \frac{SU(6)}{SO(6)}}\nn
\eeq
  \begin{tabular}{|c|c|c|c|}
    \hline
    $\GHC$ & $ (\psi, \tilde\psi)$ & $ \chi$ & Restrictions \\
    \hline\hline
    $SO(\NHC)$ & $4\times (\Sp,  \overline{\Sp})$ & $6\times \Fu$ & $\NHC = 10$\\
    \hline
    $SU(\NHC)$ & $4\times(\Fu,  \overline{\Fu})$ & $6\times \An_2$ & $\NHC = 4$\\
    \hline
  \end{tabular}}
{\footnotesize
\beq
           \frac{G_{\mathrm{F}}}{H_{\mathrm{F}}} = { \frac{SU(4)\times SU(4)'}{SU(4)_D}}{ \frac{SU(3)\times SU(3)'}{SU(3)_D}}\nn
\eeq
  \begin{tabular}{|c|c|c|c|}
    \hline
    $ \GHC$ & $ (\psi, \tilde\psi)$ & $ (\chi, \tilde\chi)$ & Restrictions \\
    \hline\hline
    $SU(\NHC)$ & $4\times (\Fu, \overline{\Fu})$ & $3\times (\An_3, \overline{\An}_3)$ & $\NHC = 7$ \\
    \hline
    $SU(\NHC)$ & $4\times (\Fu, \overline{\Fu})$ & $3\times (\An_2, \overline{\An}_2)$ & $\NHC \geq 5$ \\
    \hline
    $SU(\NHC)$ & $4\times (\Fu, \overline{\Fu})$ & $3\times (\Sy_2, \overline{\Sy}_2)$ & $\NHC \geq 5$ \\
    \hline
    $SU(\NHC)$ & $4\times (\An_2, \overline{\An}_2)$ & $3\times (\Fu, \overline{\Fu})$ & $\NHC \geq 5$ \\
    \hline
    $SU(\NHC)$ & $4\times (\Sy_2, \overline{\Sy}_2)$ & $3\times (\Fu, \overline{\Fu})$ & $\NHC \geq 8$ \\
    \hline
  \end{tabular}}
\ece
  \caption{Additional solutions giving rise to a EW coset of type $SU(4)\times SU(4)'/SU(4)_D$}\label{added}
  \end{table}

\vfill\eject

\end{document}